\documentstyle[epsf,prl,aps]{revtex}

\begin{document}
\title{ Sum rule for transport in a Luttinger liquid with long range
        interaction in the presence of an impurity.}
\author{Gianaurelio Cuniberti\\
        Dipartimento di Fisica, INFM, Universit\`a di Genova,
        Via Dodecaneso 33, I--16146 Genova, Italy}
\author{Maura Sassetti\footnote{on leave of absence 
from Istituto di Fisica di Ingegneria, INFM, Universit\`a di Genova, Italy}
 and Bernhard Kramer\\
        I. Institut f\"ur Theoretische Physik, Universit\"at Hamburg,
        Jungiusstra\ss{}e 9, D--20355 Hamburg, Germany}
\author{{\small
        (Received 14 October 1996, to be published in Europhysics Letters)}
        \vspace{3mm}}
\author{\parbox{14cm}{\parindent4mm\baselineskip11pt
        {\small
        We show that the non-linear dc transport in a Luttinger liquid with
        interaction of finite range in the presence of an impurity is
        governed by a sum rule which causes the charging energy to vanish.
        }
        \vspace{4mm} } }
\author{\parbox{14cm}{
         } }
\maketitle
\vspace{1cm}

A fundamental problem of electron transport in quantum coherent systems
is tunneling through a potential barrier \cite{landauer}. Since the
discovery that interactions play here a crucial role \cite{mooijetal} 
considerable theoretical effort has been devoted to the study of the
transport in one dimensional (1D) interacting electron systems
described by a Luttinger model \cite{luttinger,voit} with a potential
barrier \cite{fisher}.  It has been shown by renormalization group
techniques \cite{nagaosa} and by conformal field theory \cite{fendley}
that the current is suppressed at low bias voltage $U$. The
current-voltage characteristic is given by $I(U)\propto U^{2/g-1}$ 
$(U\to 0)$, with the interaction constant $g$ ($<1$ for repulsive
interaction). This implies that even an infinitesimally weak scattering
potential is completely insulating. For a tunnel junction connecting
two Luttinger systems, it has been shown that a finite range of the
interaction induces linear behavior at sufficiently high bias voltage,
$I(U)=R_{t}^{-1}(U-U_{c})$ (tunnel resistance $R_{t}$) \cite{sasskra},
with the ``charging energy'' $E_{c}\equiv eU_{c}$ proportional to the
interaction potential at zero distance. 

In this paper, we consider a potential barrier of the height $U_{b}$ in a
Luttinger liquid. The charging energy can be related to the difference
of the spectral
functions $J(\omega )$ and $J_{0}(\omega )$ of the elementary
excitations with ($g < 1$) and without ($g=1$) interaction,
respectively, via the sum rule ($\hbar=1$)
\begin{equation} E_{c}=\int _{0}^{\infty}\mbox{d}\omega
\frac{J(\omega )-J_{0}(\omega )}{\omega }
\equiv \int_{0}^{\infty}\mbox{d}\omega Z(\omega ).
\label{ec}
\end{equation}
The ``impedance function'',
$Z(\omega )$, is asymptotically
given for large and weak impurity potential 
\begin{equation}
Z(\omega )=
\left \{\begin{array}{llr}
\frac{2e^{2}}{\pi} &{\cal R}e
\left[\sigma (\omega )^{-1}-\sigma _{0}(\omega )^{-1}\right]
&\quad (U_{b}\to \infty)\\
&&\\
\frac{2\pi}{e^{2}}&{\cal R}e
\left[\sigma (\omega )-\sigma _{0}(\omega )\right]
&\quad (U_{b}\to 0)
\end{array}\right.
\label{z}
\end{equation}
with the retarded conductivities $\sigma (\omega )\equiv \sigma
(x=x',\omega)$ and $\sigma _{0}(\omega )$ of the interacting and the
non-interacting electron system, respectively, without impurity. We
will show that for arbitrary form of the interaction potential of a
finite range $E_{c}=0$ in both asymptotic limits. From this, and the
results of extensive numerical calculations,
we infer that the charging energy of an impurity in a Luttinger liquid
is always zero, independently of the hight of the barrier.  The
physical reason for this is the conservation of the total number of
excitations, as one can see from eqs.~(\ref{ec}) and (\ref{z}) and
will be discussed below in more detail.

We consider the Hamiltonian $H\equiv H_{0}+H_{b}+H_{U}$ where the
spinless Luttinger Hamiltonian $H_{0}$ corresponds to the  spectrum
$\omega _{k}=v(k)|k|$ of Bosonic pair excitations with the charge sound
velocity $v(k)=v_{F}(1+\hat{V}(k)/\pi v_{F})^{1/2}$, $v(0)\equiv
v_{F}/g$, with $\hat{V}(k)$ the Fourier transform of the interaction
potential of the range $\alpha ^{-1}$, and $v_{F}$ Fermi velocity
\cite{schulz}. The Hamiltonian of the localized potential barrier at
$x=0$, $H_{b}=U_{b}\cos{[2\pi \vartheta (x=0)]}$, is non-linear in the
phase field $\vartheta (x)$ which describes the electron density
fluctuations $\rho (x)=\rho _{0}+\partial_{x} \vartheta (x)$.

The term due to the
applied voltage is $H_{U}=e\int_{-\infty}^{\infty}dx\,U(x)\rho (x)$
and the current 
\begin{equation}
j(x)\equiv -e\dot{\vartheta} (x).
\end{equation}
Since we consider the stationary limit, the average current $I(x)$ is
independent of the position. We evaluate the current at $x=0$.
The reduced density matrix is then calculated as usual
by averaging over the bulk modes at $x\neq 0$.

The result allows to identify the dissipative kernel of the Bosonic
excitations and the effective driving force. Using imaginary time formalism
the effective Euclidean action for the ``particle'', $\vartheta (x=0,\tau
)\equiv \vartheta (\tau )$ is \cite{sassprb}
\begin{equation}
S[\vartheta ]=
\int_{0}^{\beta }\mbox{d}\tau \, \left[U_{b}\cos(2\pi \vartheta (\tau ))+
F\vartheta (\tau )\right] -\frac{1}{2}\int_{0}^{\beta }\mbox{d}\tau
\mbox{d}\tau ' \vartheta (\tau )K(\tau -\tau ')\vartheta (\tau ').
\end{equation}
Here, the external force $F\equiv e\int_{-\infty}^{\infty}dx\,E(x)\equiv U$
is independent of the spatial shape of the electric
field $E(x)=-\partial_{x}U(x)$.

The kernel $K(\tau -\tau ')$ corresponds to the inverse of the propagator
of the Bosonic excitations in imaginary time
\begin{equation}
\hat{K}(\omega _{n})=2e^{2}
\frac{\omega _{n}}{\sigma (\omega _{n})}.
\end{equation}
The conductivity without impurity potential at the Matsubara frequencies
$\omega _{n}$ is \cite{csk96}
\begin{equation}
\sigma (x,\omega_{n})=\frac{2v_{F}e^{2}}{\pi ^{2}}\int_{0}^{\infty}
\mbox{d}k\,\frac{\omega _{n}\cos(kx)}{\omega _{n}^{2}+\omega _{k}^{2}}.
\label{sigma}
\end{equation}

In order to evaluate the average current we use the real time path
integral formulation. It can be applied straightforwardly in the above
two limiting cases of strong  and weak scatterer.  For a strong
barrier, the minima of $H_{b}$ are very deep, and the variable
$\vartheta (x=0)=n$ ($n$ integer) is discrete. This
reflects that charge is transferred through the barrier only
in integer units with a tunneling probability amplitude $\Delta \equiv
\Delta (U_{b})$.

In lowest order, only paths connecting neighboring minima contribute to
the current
\begin{equation}
I(U)=\frac{\mbox{i}e\Delta ^{2}}{2}\int_{-\infty}^{\infty}\mbox{d}t\,
\sin(eUt)e^{-W(t)}
\label{current}
\end{equation}
with
\begin{equation}
W(t)=\int_{0}^{\infty}\mbox{d}\omega \,\frac{J(\omega )}{\omega ^{2}}
\left[(1-\cos{(\omega t)})\coth{\left(\frac{\beta \omega }{2}\right)}
+\mbox{i}\sin{(\omega t)}\right]
\label{w(t)}
\end{equation}
($\beta$ inverse temperature).
The spectral density $J(\omega )$ is related to the analytic
continuation of the above kernel $\hat{K}(\omega _{n})$ \cite{csk96},
\begin{equation}
J(\omega )=-\frac{1}{\pi }{\cal I}m \left[\hat{K}(-\mbox{i}
\omega +\delta )\right]
\equiv \frac{2e^{2}}{\pi }{\cal R}e \frac{\omega }{\sigma (\omega )}.
\end{equation} 

In order to avoid introducing a cutoff in eq.~(\ref{w(t)}),
we add and subtract in the exponential of eq.~(\ref{current}) 
$W_{0}(t)$, the kernel for $g=1$.
Thus, one can formally rewrite the current into a form
which resembles the semi-classical result \cite{ingold},
\begin{equation}
I(U)=\frac{1-\exp{(-\beta eU)}}{eR_{t}}
\int_{-\infty}^{\infty}\mbox{d}E \mbox{d}E'
f(E)f(-E')P(E-E'+eU)
\label{currentatt}
\end{equation}  
with $f(E)$ the Fermi func\-tion and
$R_{t}\equiv 2\omega _{max}^{2}/\pi e^{2}\Delta ^{2}\gg 2\pi /e^{2}$
the tun\-neling resi\-stance ($\omega _{max}$ cutoff frequency). The function
\begin{equation}
P(E)=\frac{1}{2\pi }\int_{-\infty}^{\infty}\mbox{d}t\,
e^{\mbox{i}Et}e^{-\left[W(t)-W_{0}(t)\right]}
\label{p(e)}
\end{equation}
has in the semi-classical theory the meaning of a probability for an
excitation with energy $E$. Here, in the microscopic model, this is no
longer the case: it can assume negative values. 

At zero temperature eq.~(\ref{currentatt}) becomes
\begin{equation}
I(U)=\frac{1}{eR_{t}}\int_{0}^{eU}\mbox{d}E\, (eU-E)P(E).
\label{currentat0}
\end{equation}
For small bias voltage we find
\begin{equation}
I(U)\propto \frac{U}{R_{t}}\left(\frac{eU}{\lambda }\right)^{2/g-2}.
\end{equation}
This result is wellknown, except that here
an intrinsic cutoff parameter $\lambda (\propto \alpha )$ appears
which reflects the decay of $Z(\omega )$ for $\omega \to \infty$ due to
the finite range of the interaction, $\alpha ^{-1}$.

In the limit of large voltage, only small times contribute to the
integral in eq.~(\ref{p(e)}). By expanding the kernel eq.~(\ref{w(t)}),
$W(t)\approx \mbox{i}t\int_{0}^{\infty}\mbox{d}\omega J(\omega )/\omega
$, and evaluating $P(E)$ from eq.~(\ref{p(e)}), one finds
$I(U)=\left(eU-E_{c}\right)/eR_{t}+\mbox{O}(1/U)$, where $E_{c}$ is
given by eq.~(\ref{ec}) and (\ref{z}) for $U_{b}\to \infty$. By
rewriting $\sigma (\omega )^{-1}-\sigma _{0}(\omega )^{-1} \equiv
-\sigma _{0}(\omega )^{-1}\sum_{n=0}^{\infty} [\sigma (\omega )/\sigma
_{0}(\omega )-1]^{n+1}$, and using that $\lim_{\omega_{max}\to
\infty}\sigma _{0}(\omega )=\mbox{const.}$, one can straightforwardly
show via Fourier transformation that ${\cal R}\mbox{e}\int
_{0}^{\infty}\mbox{d}\omega [\sigma (\omega )-\sigma _{0}(\omega
)]^{n}=0$, since $\hat{\sigma }(t),\hat{\sigma }_{0}(t)\propto \Theta
(t)$ ($\Theta $ Heavyside function) \cite{cuniberti}. This implies
$E_{c}=0$. 

\begin{figure}[htb]
\label{currentvoltage}
      \epsfysize=9.truecm
      \epsfbox{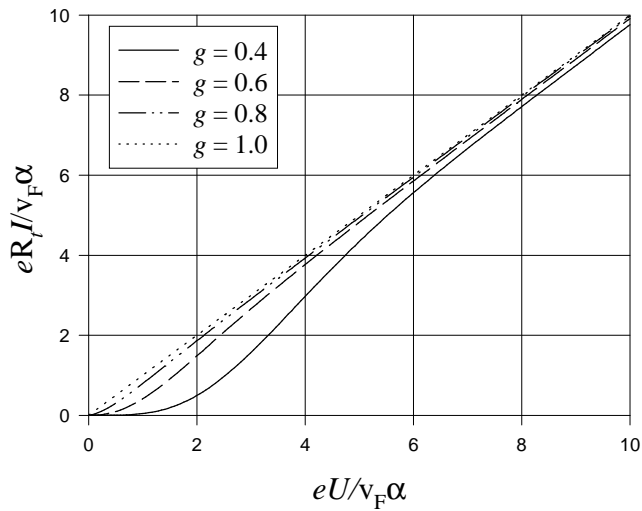}
\caption[1]{The current-voltage characteristic of a high potential
barrier in a Luttinger liquid with an interaction potential of finite
range $\alpha ^{-1}$ for different strengths of the interaction, $g^{-2}$.}
\end{figure}

We evaluated the $I(U)$ for various
forms of the interaction potentials, which were obtained by projecting
a 3D screened Coulomb repulsion to a quasi-1D quantum wire \cite{csk96}.
We found that the charging energy always vanishes for high barrier within the
numerical errors ($<10^{-9}$), as long as the range of the interaction
is finite.
Figure~\ref{currentvoltage} shows the result for the
``Luttinger limit''
$
V(x)=V_{0}\alpha e^{-\alpha |x|}.
$
It is defined by the requirement that the screening length $\alpha
^{-1}$ is much smaller then the effective width of the wire.
The non-analytical small-voltage behavior which depends strongly on the
interaction constant $g\equiv v_{F}/v(0)$ is clearly depicted. For
large voltage all of the curves tend asymptotically to the
non-interacting limit, $I(U)=U/R_{t}$, in accordance with the above
analytic finding.
It is important to notice that for the vanishing of the charging energy
it is necessary to have a point of inflection in $I(U)$. The existence
of the latter is guaranteed by the fact that the asymptotic limits of
$J(\omega )$ for small and large frequency behave as 
\begin{equation}
J(\omega )=2\omega 
\left\{
\begin{array}{lr}
\frac{1}{g}+\mbox{O}\left(\frac{\omega }{\alpha }\right)^{2}
			\quad&(\omega \to 0)\\
	\\
1-\frac{V_{\omega }}{2\pi }\quad &(\omega \to \infty)	
\end{array}\right.
\label{asymp}
\end{equation}
Here, $V_{\omega }\equiv \hat{V}(\omega /v_{F})$.

For small potential barrier, the current can be evaluated by using the
wellknown duality argument \cite{schmid,zwerger,sassetti}. The 
calculation to second order gives
\begin{equation}
I(U)=G_{0}U-\frac{\mbox{i}}{2}geU_{b}^{2}
\int_{-\infty}^{\infty}\mbox{d}t\,\sin{(geUt)}\,e^{-\tilde{W}(t)}
\label{weakbarrieri(u)}
\end{equation}
with the renormalized dc conductance of the Luttinger wire,
$G_{0}=ge^{2}/2\pi $. The function $\tilde{W}(t)$ has the same form as
$W(t)$, however with a spectral density $\tilde{J}(\omega )$ that is
related to $J(\omega )$ by the duality transformation \cite{sassetti}
\begin{equation}
\tilde{J}(\omega )=\frac{(2\pi \omega )^{2}J(\omega )}{|K(\omega )|^{2}}=
\frac{2\pi }{e^{2}}\omega {\cal R}e \sigma (\omega ).
\end{equation}
At $T=0$ one obtains now
\begin{equation}
I(U)=G_{0}U-\frac{1}{eR_{b}}\int_{0}^{eU}dE\,(eU-E)\tilde{P}(E) 
\end{equation}
where $\tilde{P}(E)$ is given again by eq.~(\ref{p(e)}),
but with $\tilde{W}(t)$
instead of $W(t)$ and with the resistance of barrier given by
$R_{b}=2\omega _{max}^{2}/\pi e^{2}gU_{b}^{2}$.

At sufficiently high voltage, the current becomes now
\begin{equation}
I(U)=(G_{0}-R_{b}^{-1})U+\frac{E_{c}}{eR_{b}},
\end{equation}
and always
\begin{equation}
E_{c}=\frac{2\pi }{e^{2}}{\cal R}e\int_{0}^{\infty}\mbox{d}\omega \,
(\sigma (\omega )-\sigma _{0}(\omega ))=0.
\end{equation}
The latter equality can easily been seen by evaluation of the frequency
integral using eq.~(\ref{sigma}). One observes that that the vanishing
of the charging energy is nothing but a consequence of the conservation
of the number of elementary excitations, independently of the form of
the interaction.
  
For small bias, but $eU/\lambda \gg
(U_{b}/\lambda )^{1/2(1-g)}$, one finds
\begin{equation}
I(U)-G_{0}U\approx\frac{U}{R_{b}}\left(\frac{eU}{\lambda }\right)^{2(g-1)}.
\end{equation}

When assuming that for very small voltage
the correct result was, as described above, $I(U)\propto
U^{2/g-1}$ \cite{fisher}, a point of inflection must exist also for
small barrier in the current-voltage curve. The point of inflection
indicates the crossover from the suppression of the current due to the
coupling to the bath of bulk modes of the Luttinger wire to the
behavior at high voltage where interaction becomes unimportant, in this
model. 

From the asymptotic results for very high and small potential barrier
that have been confirmed for several different forms and strengths of
the interaction potential by numerical evaluation \cite{cuniberti}, we
conclude that any potential barrier in a Luttinger model of 1D
interacting electrons gives always zero charging energy, as long as the
interaction potential is of finite range. 

This is also suggested by results obtained for the model of a tunnel junction
connecting two semi infinite Luttinger systems \cite{cuniberti}. Here,
we found
$E_{c}=0$ if the strength of the interaction
between electrons located at different sides of the junction, $V_{12}$,
is the same as for electrons on the same side, $V_{11}$. A non-zero
charging energy is however obtained when $V_{12}/V_{11}<1$.
Generalization of the present scattering model to such an interaction
potential yields the same result. This shows also that the approximations
involved in both models, namely using a tunneling Hamiltonian in the
former, and the ``instantaneous jump approximation'' when evaluating
the path integral in the latter, are equivalent.

In the tunnel junction model, the physics is more clearly
displayed: the {\em absence} of the charging effect at higher voltage
is related to the {\em presence} of the interaction between electrons
left and right of the barrier. It is {\em only} for zero range
interaction that the current persists to be suppressed up to the bias
voltage which corresponds to the cutoff frequency $U_{max}=\omega
_{max}/e$. Any finite interaction range will eventually, for
sufficiently large bias voltage (but smaller than the cutoff voltage),
yield the crossover to Ohm's law $I(U)=U/R$. As a consequence,  the
analogue to the classical ``capacitance'' of the system described by
the present microscopic model diverges.

Generalizing the model of a tunnel junction between two Luttinger
systems with zero range interaction (with cutoff) to include many ($N$)
transport channels \cite{matveev} yielded $g(N)\to 1$ for $N\to
\infty$, such that the suppression of the current for small bias 
vanishes in this limit. Since the interaction becomes small with
increasing $N$, we  expect that our above, somewhat intriguing result
will remain essentially unchanged when including the influence of many
channels.

This should render the ``capacitance of a tunnel junction''
proportional to the area of the junction ($\propto N$) to be {\em 
unobservable} as long as the interaction between electrons on different
sides is assumed the same as between those on the same side of the junction.   

It is indeed wellknown that it is very difficult to observe
experimentally Coulomb blockade using a single tunnel junction due to
the presence of a large shunt capacitance \cite{delft,delsing}. Our
above result offers an explanation for the {\em microscopic} source of
this shunt capacitance: the interaction between electrons left and
right of the impurity. 

Discussions with Det\-lef Heit\-mann and Wolf\-gang Han\-sen are
gra\-te\-ful\-ly ack\-now\-led\-ged. This work has been supported by
the EU via HCM and TMR prog\-rammes, contracts CHRX-CT93 0136,
CHRX-CT94 0464 and FMRX-CT96 0042, re\-spec\-ti\-vely, and by the
Deut\-sche For\-schungs\-ge\-mein\-schaft via the
Gra\-du\-ier\-ten\-kol\-leg ``Physik nano\-struk\-tu\-rier\-ter
Fest\-k\"or\-per'' of the Uni\-ver\-si\-t\"at Ham\-burg.

\end{document}